\documentclass[12pt]{article}
\usepackage{amssymb,amsmath,epsfig}
\usepackage{graphicx}
\usepackage{xcolor}
\usepackage{dsfont}
\allowdisplaybreaks
\begin{document}
\title{\textbf{Bulk Viscosity and Thermodynamic Approaches to
Singularity Resolution in Non-Metric Gravity}}
\author{M. Sharif$^{1,2}$\thanks{msharif.math@pu.edu.pk} ,
M. Zeeshan Gul$^{3,4,1}$\thanks{mzeeshangul.math@gmail.com} ~and
Nusrat Fatima$^1$\thanks{nusratfatimaliaqat@gmail.com}\\
$^1$ Department of Mathematics and Statistics, The University of Lahore,\\
1-KM Defence Road Lahore-54000, Pakistan.\\
$^2$ Research Center of Astrophysics and Cosmology, Khazar University,\\
Baku, AZ1096, 41 Mehseti Street, Azerbaijan.\\
$^3$ College of Transportation, Tongji University, Shanghai 201804,
China.\\
$^4$ Postdoctoral Station of Mechanical Engineering, Tongji
University, \\ Shanghai 201804, China.}

\date{}
\maketitle

\begin{abstract}
This manuscript investigates the impact of bulk viscosity on the
viability of cosmic bounce solutions in the context of
$f(\mathcal{Q})$ theory, where $\mathcal{Q}$ is a non-metricity
scalar. To complete this objective, we study the behavior of an
isotropic homogeneous universe filled with a perfect fluid and
consider a innovative parametrization of bulk viscosity coefficient
with an arbitrary constant $\zeta_0$ as $\zeta=\zeta_{0}H$. We
consider the particular functional form of this modified theory to
explore how this gravitational framework influences the cosmic
evolution and explore a range of cosmological parameters to assess
the existence and physical viability of bounce solutions. We also
investigate the evolution of entropy through the second law of
thermodynamics. The positive trend of energy density, negative
pressure profile and violation of energy conditions support the
existence of viable cosmological bounce scenario and highlights the
significance of bulk viscosity in this framework. These results
demonstrate that $f(\mathcal{Q})$ gravity provides a compelling
alternative to standard cosmic models and provides deep insights
into the nature of gravitational interaction and the early cosmos.
\end{abstract}

\section{Introduction}

Observational data from Type-Ia redshift Supernovae surveys
\cite{1}, measurements by Wilkinson microwave anisotropy probe
\cite{2} and cosmic microwave background \cite{3} strongly support
the cosmic expansion. Researchers put forward that this cosmic
expansion is driven by a mysterious entity, generally referred to as
dark energy (DE), which is thought to exert a large negative
pressure. This mysterious energy has encouraged the scientists to
discover its hidden aspects. To account the effects of DE, Einstein
added cosmological constant into his field equations , which is
widely known as $\Lambda$CDM model. However, this model undergo
coincidence problem and the fine-tuning issue. These challenges
motivated the research community to modify general theory of
relativity (GR) to gain deeper insights for DE and cosmic
acceleration \cite{4}. Such modifications reshape the geometrical
structure of the Einstein-Hilbert formulation, thereby providing
modified gravitational models to resolve the cosmic mysteries. The
dynamical substitute based on Hamiltonian analysis is calculated in
\cite{4a}.

Within the context of GR, gravitational phenomena can be represented
through three distinct yet equivalent geometric representations. The
first based on the spacetime's curvature with underlying the
assumptions of zero torsion and non-metricity. The second represents
the teleparallel approach, which remove the curvature but reserves
torsion as the basic geometric entity. In this context, the
gravitational field is given through tetrad fields instead of metric
tensor, thereby incorporating torsional effects. This model
describes the gravity through torsion-based interactions instead of
curvature. A non-metricity-based description, which produces a
symmetric teleparallel gravity, is a third alternative but
equivalent representation \cite{12}. The main objective is to
establish a more inclusive theoretical model to understand the
gravity and cosmology. By dealing the current observational
discrepancies, it seeks to offer profound insights into the
essential features of gravity and the cosmos, clarifying the
enquiries regarding cosmic structure and nature. Detailed analysis
of non-metricity based framework is explored in \cite{z1}-\cite{z6},
while observational constraints in modified theories are discussed
in \cite{MT1}-\cite{z20}.

Observational evidence claims that the cosmos originated from a
singularity, an extremely dense and hot state where all matter was
compressed, commonly referred to as the big bang. Despite of its
advantages, this theory still faces some cosmic limitation
\cite{c028}. One major issue of this framework is its inability to
explain the origin of the initial singularity or the preceding
circumstances, leaving the universe's initial state unclear.
Furthermore, the big bang scenario finds it crucial to justify some
observational evidences, like galaxy distribution,  leading to
inflation theory \cite{c029}. This theory provides a significant
basis for comprehending the cosmic expansion and solves key
issues,yet the singularity issue remains unaddressed. Bouncing
cosmology has emerged as a promising alternative to prevent the
initial singularity in theoretical cosmic models \cite{c030}. In
accordance to the cosmic bounce model, the universe follows a cyclic
pattern of expansions and contractions instead originating from a
single big bang event. The weak equivalence principle for a
Schwarzschild gravitational field based on the light-clock model has
been discussed in \cite{PD1}. The dynamical substitute based on
Hamiltonian analysis has been examined in \cite{PD2}.

The idea of bouncing cosmology in the context of alternative gravity
theories has attracted greater importance from researchers due to
its intriguing properties. Bajardi et al \cite{25} examined the
cosmological bouncing models in $f(\mathcal{Q})$ gravity, caculating
the cosmic wave function via Hamiltonian approaches. Mandal et al
\cite{26} evaluated the feasibiity of matter-bounce phenomena using
distinct $f(\mathcal{Q})$ functional forms. Malik and Shamir
\cite{27} explored the non-singular bouncing solutions to study
cosmic dynamics in $f(\mathcal{R})$ theory. Ilyas et al \cite{28}
explored the dynamical behavior of cosmological bounce solutions in
the same theory. Lohakare et al \cite{30} studied the emergenve of a
cosmic bounce in $f(\mathcal{G})$ gravity ($\mathcal{G}$ is the
Gauss-Bonnet invariant) using energy constraints. Bozza and Burni
\cite{c032} examined the anisotropic cosmic bounce. Cai et al
\cite{c033} give ride  a non-singular bouncing cosmology model using
a scalar field. This approach offers a more viable framework for
comprehending the cosmic early accelerated. Bamba et al \cite{36}
used several scale factor forms to study $f(\mathcal{R})$ gravity
and found feasible bouncing solutions. Amani \cite{38} used the
redshift variable to investigate the bouncing solution in the same
gravity. Tripathy et al \cite{39} studied the bouncing solutions of
the extended theory of gravity.

This study investigates the feasibility of achieving a bouncing
cosmology in $f(\mathcal{Q})$ gravity with a bulk viscous fluid in
an Friedmann-Robertson-Walker (FRW) background. We analyze how the
interplay between $f(\mathcal{Q})$ gravity and bulk viscosity can
generate a non-singular bounce model. The organization of this paper
is as follows. Section \textbf{2} evaluates the bounce dynamics and
solves the gravitational field equations using bulk viscosity
coefficient. Section \textbf{3} examines the behavior of kinematic
parameters such as the scale factor, Hubble parameter and
deceleration parameter, the equation of state (EoS) parameter and
behavior of energy conditions (ECs) through graphical analysis. The
brief analysis of the second law of thermodynamics is provided in
section \textbf{4}. Finally, section \textbf{5} summarizes our
findings.

\section{$f(\mathcal{Q})$ Gravity: Field Equations}

The corresponding integral action is defined as \cite{88}
\begin{equation}\label{1}
S=\int\frac{1}{2}f(\mathcal{Q})\sqrt{-g}d^{4}x+\int
L_M\sqrt{-g}d^{4}x,
\end{equation}
where $L_{M}$ is the matter-Lagrangian density and $g$ is the
determinant of the metric tensor. The non-metricity is given by
\begin{equation}\label{2}
\mathcal{Q}=-\mathcal{Q}_{\xi\alpha\beta}P
^{\xi\alpha\beta}=-\frac{1}{4} (-\mathcal{Q}^{\xi\alpha\beta}
\mathcal{Q}_{\xi\alpha\beta}+2 \mathcal{Q}^{\xi\alpha\beta}
\mathcal{Q}_{\beta\xi\alpha} -2
\mathcal{Q}^{\xi}\tilde{\mathcal{Q}}_{\xi}+\mathcal{Q}^{\xi}\mathcal{Q}_{\xi}),
\end{equation}
where $P^{\xi\alpha\beta}$ is the superpotential is expressed as
\begin{equation}\label{3}
P^{\xi}_{\alpha\beta}=-\frac{1}{2}L^{\xi}_{\alpha\beta}
+\frac{1}{4}(\mathcal{\mathcal{Q}}^{\xi}
-\tilde{\mathcal{Q}}^{\xi})g_{\alpha\beta}- \frac{1}{4} \delta
^{\xi}_{[\alpha \mathcal{Q}_{\beta}]}.
\end{equation}
Here, the disformation tensor $(L^{\xi}_{~\alpha\beta})$ describes
the overall expansion or contraction of spacetime, defined as
\begin{equation}\label{4}
L^{\xi}_{~\alpha\beta}=-\frac{1}{2}g^{\xi\lambda}\big(\nabla_{\beta}g_{\alpha\lambda}+
\nabla_{\alpha}g_{\lambda\beta}-\nabla_{\lambda}g_{\alpha\beta}\big),
\end{equation}
and
\begin{equation}\label{5}
\mathcal{Q}^{\xi}=\mathcal{Q}^{\xi~~\alpha}_{~\alpha} ,\quad
\tilde{\mathcal{Q}}^{\xi}= \mathcal{Q}^{\alpha}_{~~\xi\alpha}.
\end{equation}
The corresponding field equations are
\begin{equation}\label{7}
\frac{2}{\sqrt{-g}}\nabla_{\xi}(f_{\mathcal{Q}}\sqrt{-g}
P^{\xi}_{~\alpha\beta})+\frac{1}{2}f(\mathcal{Q})
g_{\alpha\beta}+f_{\mathcal{Q}}
(P_{\alpha\xi\tau}\mathcal{Q}_{\beta}^{~\alpha\xi}-
2\mathcal{Q}^{\xi\tau}_{~~~\alpha}
P_{\xi\tau\beta})=T_{\alpha\beta},
\end{equation}
where $f_\mathcal{Q}\equiv\frac{\partial f}{\partial \mathcal{Q}}$.

To investigate the unknown aspects of the universe, we assume FRW
spacetime with scale factor $a(t)$ as
\begin{equation}\label{9}
d{s}^2=-d{t}^2+(d{x}^2+d{y}^2+d{z}^2)a^{2}(t).
\end{equation}
This homogeneous and isotropic universe model provides a valuable
framework to explore various cosmological phenomena, offering
insights into the behavior of the cosmos. To achieve a more
realistic depiction of cosmic fluid, we consider the existence of
bulk viscosity in the isotropic fluid as
\begin{eqnarray}\label{11}
T_{\alpha\beta}=(\rho+\mathcal{P})U_{\alpha}U_{\beta}+\tilde{\mathcal{P}}
g_{\alpha\beta}, \quad \tilde{\mathcal{P}}=\mathcal{P}-3{H}\zeta,
\end{eqnarray}
where $\zeta$ represents bulk viscosity coefficient and $H$ is the
Hubble parameter.

In bouncing cosmology, the presence of a viscous fluid is vital to
account for dissipative processes that are not present in a perfect
fluid.  The existence of viscous effect alters the effective
pressure and allows a gradual entrance into the expanding phase. A
positive bulk viscosity coefficient is a dissipative process that
helps the universe intrinsic entropy production and can produce an
effective negative pressure. This adverse pressure could offset
gravitational collapse and can possibly resemble dark energy-like
behavior. The positive viscosity is significant in enabling a cosmic
bounce, as it violates the effective null energy bound without
requiring any exotic stuff.  When $\zeta=0$, the fluid simplifies to
a normal perfect fluid, and no entropy generation owing to
dissipative effects. This scenario allows us to separate the
contributions from the geometric modifications and separate them
from those emerging owing to viscosity.  A negative $\zeta$
corresponds to entropy drop, which leads to instabilities in cosmic
evolution. Therefore, while theoretically possible, such scenarios
are thermodynamically and physically disfavored.

Using Eqs.\eqref{7}-\eqref{11}, we obtain
\begin{eqnarray}\label{13}
3{H}^{2}&=&\frac{1}{2f_{\mathcal{Q}}}\bigg(-\rho+\frac{f}{2}\bigg),
\\\label{14}
2\dot{H}+3H^{2}+f_{\mathcal{Q}\mathcal{Q}}H&=&\frac{1}{2f_{\mathcal{Q}}}
\bigg({\tilde{\mathcal{P}}}+\frac{f}{2}\bigg).
\end{eqnarray}
To study the universe behavior, we take the functional form as
\cite{93}
\begin{equation}\label{15}
f(\mathcal{Q})=\psi \mathcal{Q},\quad\psi\neq0,1.
\end{equation}
This choice exhibits nontrivial features due to the dynamical
interplay between bulk viscosity and the modified gravity background
which is absent in GR. This cosmic model captures considerable
applications in explaining different cosmic phenomena, specifically
describing the accelerated cosmic expansion without requiring for
exotic DE components. Within the framework of our model setting, the
incorporation of bulk viscosity provides an approach for realizing
non-singular cosmic bounce, affirming the thermodynamic
compatibility during cosmic modulations. This model can further be
used to study the structure formation, late-time acceleration and
inflationary dynamic. This model is inspired by several thoughts in
the context of alternative gravity theories \cite{94}-\cite{96}.

Using Eqs.\eqref{13}-\eqref{15}, we have
\begin{equation}\label{18}
-2\psi{\dot{H}}+(\rho+\mathcal{P})-3{H}\zeta=0.
\end{equation}
To obtain the exact solution of the field equations \eqref{13} and
\eqref{14}, two additional equations are required as the field
equations involve four unknown parameters ($\rho,~
\tilde{\mathcal{P}},~\zeta$ and $H$). In this setting, we assume the
following expression for pressure and density as
\begin{eqnarray}\label{19}
\rho=\frac{\mathcal{P}}{(1-\mu)}, \quad 0<\mu<1.
\end{eqnarray}
Additionally, we assume the bulk viscosity coefficient as
\begin{eqnarray}\label{20}
\zeta={\zeta}_{0}H.
\end{eqnarray}
Using Eq.\eqref{20}, the expression of energy density becomes
\begin{equation}\label{21}
\rho=\frac{3({\zeta}_{0}+\psi)H^2+2 \psi\dot{H}}{(-1+\mu)}.
\end{equation}
Using Eqs.\eqref{18}-\eqref{21}, we have
\begin{equation}\label{22}
-3{\zeta}_{0}H^{2}+3({\zeta}_{0}+\psi)H^{2}+\frac{3({\zeta}_{0}+\psi)H^2+2
\psi\dot{H}}{(-1+\mu)}=0,
\end{equation}
whose solution yields
\begin{equation}\label{23}
H=\frac{2\psi}{3\zeta_{0}t+3\mu\psi t-2\psi{{\Phi}_{1}}}.
\end{equation}
The relation $H=\frac{\dot{a}}{a}$ gives
\begin{equation}\label{24}
a(t)=\Phi_{2}\big[3t(\zeta_{0}+\mu\psi)H^{2}-2\psi\Phi_{1}\big]^{\frac{2}{3}\frac{\psi}
{(\zeta_{0}+\mu\psi)}},
\end{equation}
where $\Phi_{1}$ and $\Phi_{2}$ are integrating constants.

\section{The Bouncing Universe}

According to bouncing cosmology, the universe undergoes phases of
expansion and contraction to avoid the singularity. According to
this theory, quantum events prevent the universe from collapsing
when it reduces to its smallest size, causing a bounce that
initiates a new expanding phase. Bouncing cosmology suggests a
cyclic sequence of contraction to expansion, in contrast to
cosmological frameworks that postulate a primal singularity. This
phenomena offers new insights on early-cosmic behaviour in addition
to redefining the origins of the cosmos.  In the next subsections,
we analyze the behavior of several cosmic parameters for realistic
bouncing cosmology.

\subsection{Analysis of Hubble Parameter}

The Hubble parameter is essential to understanding cosmic behaviour
as it measures the universe's expansion and contraction phases. The
Hubble parameter is positively correlated with cosmic expansion and
negatively correlated with contraction. The Hubble parameter
indicates the point at which the contracting and expanding phases
interact at the bounce epoch ($H=0$). Figure \textbf{1} demonstrates
that the Hubble parameter transitions from negative to positive
phase near the bounce point with $\dot{H}=4G\rho(1+\omega)>0$,
showing the cosmic expansion before transitioning into contraction
phase for negative values of $\psi$. The Hubble parameter disappears
exactly at the bounce epoch. Additionally, the $\zeta_0$ values are
bound to $\zeta_0 \leq 10.13$, guaranteeing that the model stays
valid inside parametric bonds.
\begin{figure}
\epsfig{file=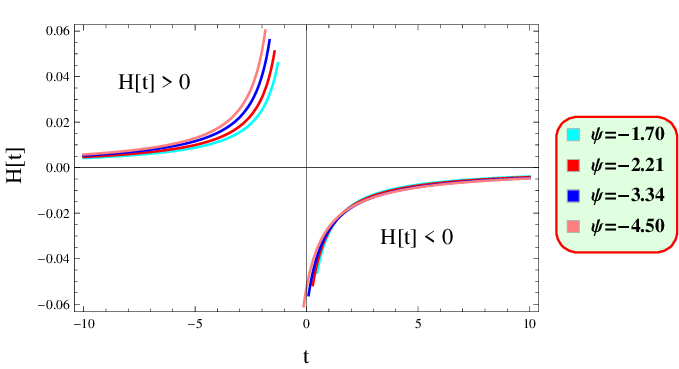,width=.5\linewidth}
\epsfig{file=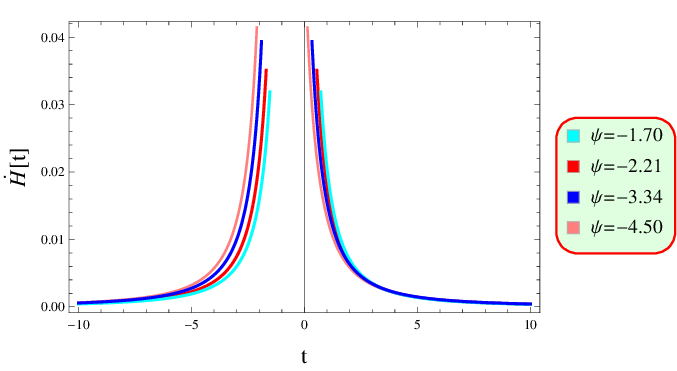,width=.5\linewidth}\caption{Plots of Hubble
parameter to cosmic time.}
\end{figure}

\subsection{Investigation of Deceleration Parameter}

One essential cosmic quantity that characterizes the rate of cosmic
expansion is the deceleration parameter. It ascertains how the
universe's expansion changes over time, indicating whether the
expansion period is accelerated or decelerated.  The value of the
deceleration variable is
\begin{equation}\label{27}
q=\frac{3}{2}\bigg(\mu+\frac{\zeta_{0}}{\psi}\bigg)-1.
\end{equation}
This parameter helps us to comprehend whether the cosmic expansion
is slowing down or speeding up. Positive values of $q$ indicate that
the expansion is slowing down due to the gravitational attraction of
matter, known as deceleration. On the other hand, negative values of
$q$ determine the accelerated cosmic expansion, suggesting the
influence of a mysterious energy phenomenon with negative pressure,
i.e., DE. A value of $q=0$ corresponds to a universe that is
expanding at a constant rate, that is a coasting universe with zero
acceleration and zero deceleration. This case represents a critical
boundary between decelerating and accelerating phases, and can
appear in various cosmological models. Table \textbf{1} shows that
the classification of cosmic models relies on how the Hubble and
deceleration variables vary over time. According to this
classification, cases \textbf{IV} and \textbf{V} are feasible, given
that the current phenomenon depicts an expanding universe \cite{98}.
\begin{table}\caption{Cosmic classification using deceleration and
Hubble parameters.}
\begin{center}
\begin{tabular}{|c|c|c|c|}
\hline Case & Condition & Universe Nature
\\
\hline I & $q>0$, $H>0$ & Deceleration and Expansion
\\
II & $q>0$, $H<0$ & Deceleration and Contraction
\\
III & $q>0$, $H=0$ & Deceleration and Bounce Point
\\
IV & $q<0$, $H>0$ &   Acceleration and Expansion
\\
V  & $q<0$, $H<0$ & Acceleration and Contraction
\\
VI & $q<0$, $H=0$ & Acceleration and Bounce
\\
VII & $H=0$, $q=0$ & Static
\\
\hline
\end{tabular}
\end{center}
\end{table}

Figure \textbf{2} represents that the cosmos is undergoing constant
accelerated expansion. In this scenario, the rate of expansion is
not only sustained but also increases over time, indicating that the
effect of repulsive forces (DE) dominates over gravitational
deceleration. This accelerated trend is significant in explaining
the late-time cosmic expansion, as observed in cosmological data.
\begin{figure}\center
\epsfig{file=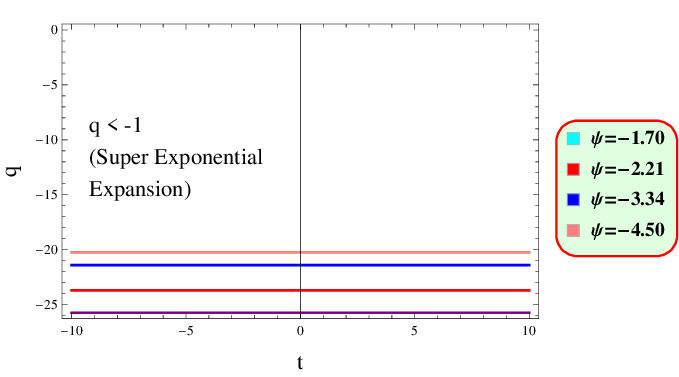,width=.5\linewidth}\caption{Plots of
deceleration parameter to cosmic time.}
\end{figure}

\subsection{Study of Matter Variables}

Through matter variables, we investigate how the bulk viscosity
fluid configuration affects the cosmic geometric structure.  The
pressure dynamic and energy density evolution over time are shown in
Figure \textbf{3}. Stability with the cosmic matter configuration is
ensured by the positive trend of energy density reaching its maximum
value and staying bounded throughout the cosmic behavior.
Conversely, a propelling factor that controls the cosmic rapid
expansion is indicated by the negative profile. The DE or other
matter components that cause the rapid expansion are characterized
by this negative pressure behavior.
\begin{figure}
\epsfig{file=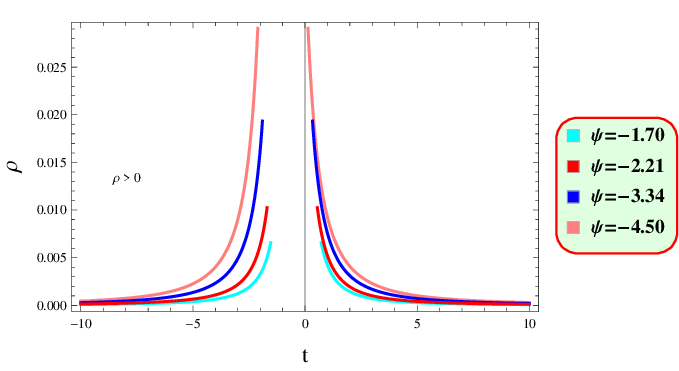,width=.5\linewidth}
\epsfig{file=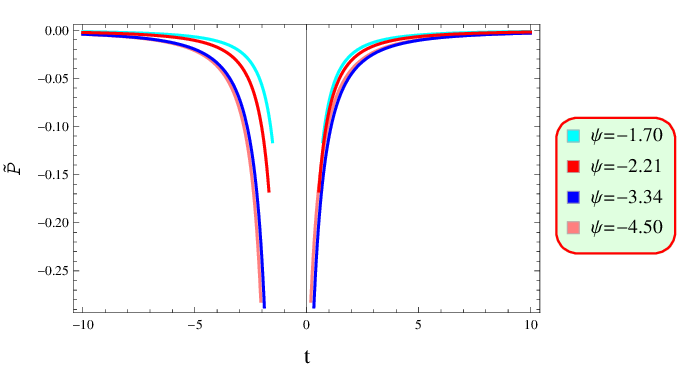,width=.5\linewidth}\caption{Plots of energy
density and pressure.}
\end{figure}

\subsection{Analysis of State Parameter}

In cosmology, the EoS parameter $(\omega)$ characterizes the
relationship between the pressure and energy density which is
essential to comprehend the evolution of the universe under
different cosmological models. The dynamics of EoS parameter is
important in bouncing cosmology, as it determines how the universe
evolves during the compression, bounce and expansion epochs. The
value of $\omega$ impacts whether the universe can successfully
undergo a bounce, avoid singularities and transition smoothly
between these phases. The EoS parameter is classified into distinct
phases of cosmic evolution. The matter-dominated regions such as
dust, radiative fluid and stiff matter regions are determined by
$\omega=0,\frac{1}{3}$, $1$, respectively. Whereas the vacuum,
phantom and quintessence cosmic phases are represented by
$\omega=-1$, $\omega < -1$, $-\frac{1}{3}<\omega<-1$, respectively
\cite{99}. For the viscous fluid in $f(\mathcal{Q})$ gravity model,
the EoS parameter is found to be
\begin{equation}\label{28}
\omega=-\frac{6[\zeta_0 + (-1+\mu)\psi](\zeta_0+\mu \psi)}{\psi
[3t(\zeta_0+\mu\psi)-2\psi\Phi_1]}.
\end{equation}
\begin{figure}\center
\epsfig{file=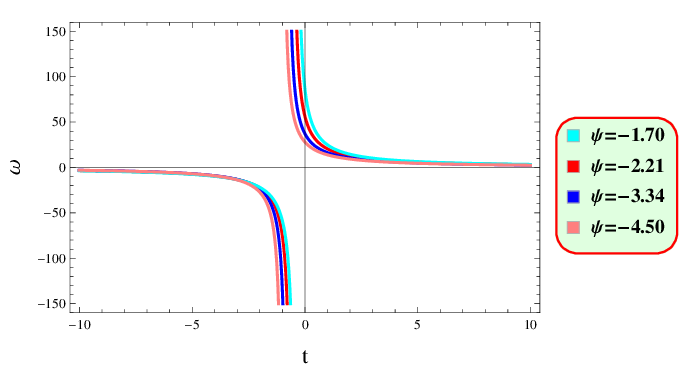,width=.5\linewidth}\caption{Plots of EoS
parameter versus cosmic time.}
\end{figure}

The variation of the EoS parameter is depicted in Figure \textbf{4}.
Before the bounce $(t<0)$, the EoS parameter remains negative and
decreases with time which indicates a phase dominated by an exotic
matter component. The incorporation of bulk viscosity improves this
effect by inserting dissipative pressure, efficiently increasing the
contribution of negative pressure and confirms a smooth approach to
the cosmic bounce. At the bouncing spot $(t=0)$, the state parameter
experiences a transition as the cosmos approaches its minimum scale
factor. After the subsequent bounce $(t>0)$, the EoS parameter shows
the positive profile and eventually diverges to zero. This
incremental decay towards zero illustrates that the universe
convergently approaches to a radiation dominated or dust-like
configuration expansion at late epochs.
\subsection{Cosmic Energy Constraints}
\begin{figure}
\epsfig{file=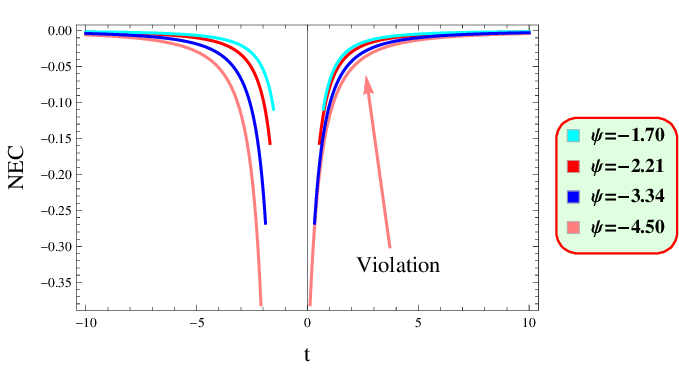,width=.5\linewidth}
\epsfig{file=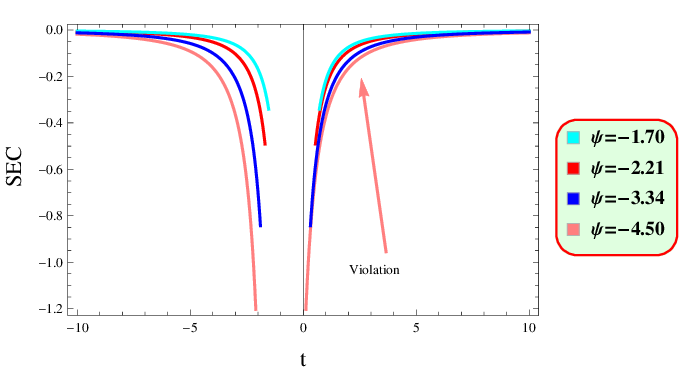,width=.5\linewidth}
\epsfig{file=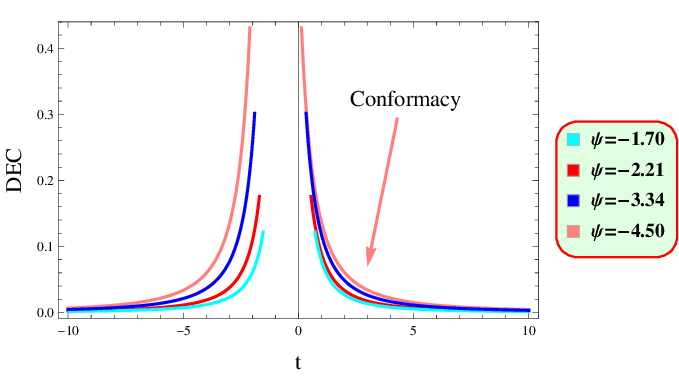,width=.5\linewidth}
\epsfig{file=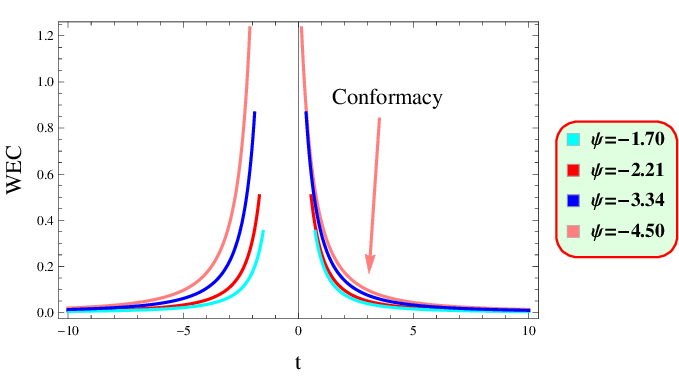,width=.5\linewidth}\caption{Plots of energy
conditions corresponding to cosmic time.}
\end{figure}

The ECs play a major role in gravitational physics, as they offer
the foundation for the presence of black holes and spacetime initial
singularities \cite{100}. To assess the feasibility of the cosmic
model in the framework of accelerated expansion, we analyze the ECs
in this section. These ECs are deivided into four categories such as
null energy condition (NEC: $\rho+\tilde{P}\geq0$ and $\rho\geq0$),
weak energy conditions (WEC: $\rho+\tilde{P}\geq0$), strong energy
conditions (SEC: $\rho+3\tilde{P}\geq0$ and $\rho+\tilde{P}\geq0$)
and dominant energy conditions (DEC: $\rho\geq|\tilde{P}|$),
\cite{101}. By utilizing these constraints, researchers investigate
the viability for several cosmic structures. The inclusion of a bulk
viscosity parameter in this study significantly impacts the ECs
within the universe.

Figure \textbf{5} depicts the evolution of each ECs corresponding to
cosmic time. The violation of NEC and SEC lead to a decrease in
energy density and thus cause the cosmic expansion. For a bounce to
take place, it is necessary that the NEC and SEC are depleted at the
bounce epoch \cite{101}. Moreover, according to the most recent
observational data on the expanding cosmos, there must be a
violation of SEC. The universe rapid expansion is demonstrated by
the behavior of SEC. Compared to certain other models, this analysis
shows that the WEC is satisfied, indicating that the matter
distribution remains physically realistic and consistent with a
perfect fluid model. This compliances with the WEC and fulfillment
of the DEC reinforces the argument for a non-singular viable
bouncing universe, free from the initial singularity issue.

\section{Second Law of Thermodynamics}

The universe dynamics lies on the generalized second law of
thermodynamics, which demands the total entropy production of a
closed system do not decreases but remains constant or grows overs
time. Study of entropy production behavior is the crucial aspect in
bouncing cosmology throughout the contraction and expansion periods.
Moreover, the rise in Hawking temperature highlights that the system
progresses towards the equilibrium during cosmic expansion. This
increment leads to theoretical observation of reheating phenomena
following a cosmological bounce. The collaboration between the
Hawking temperature behavior and entropy production signifies the
leading role of bulk viscosity in sustaining a thermodynamically
viable cosmic bounce. The smooth thermodynamic modulations support
the viability of the model in resolving cosmic singularities while
adhering to fundamental principles of thermodynamic. The total
cosmic entropy incorporates the contributions from both the matter
within the horizon ($S_{\text{in}}$) and the entropy associated with
its boundary ($S_{\text{on}}$).

In this analysis, the boundary of the universe is determined by the
apparent horizon, whose radius is expressed in terms of the Hubble
variable \eqref{23} as
\begin{equation}\label{32}
R_H=\frac{1}{H}=\frac{-2\Phi_{1}\psi+3\zeta _{0}t+3\mu
t\psi}{2\psi}.
\end{equation}
The horizon surface area is related to the entropy, which can be
expressed as
\begin{equation}\label{33}
S_{\text{on}}=\frac{K_{b}A}{4{l}^{2}_{p}},
\end{equation}
where $K_b$ is the Boltzmann constant, $l_p$ represents the Planck
length and $A=4\pi R_{H}^{2}$ denotes the area of the cosmic
horizon. Now, using Eq.\eqref{32} and \eqref{33}, we have
\begin{equation}\label{34}
\dot{S}_{\text{on}}=\frac{K_b\pi\big(3\zeta _{0}+3\mu \psi\big)
\big(3\zeta_{0}t+3\mu t\psi-2\psi\Phi_{1}\big)}{2\psi^2{l}^{2}_{p}}.
\end{equation}
The entropy of matter enclosed by the horizon, as determined by the
Gibb's relation is
\begin{equation}\label{35}
T_{H}dS_{\text{in}}=d({\rho}V)+{p}dV.
\end{equation}
Here, the Hawking temperature $(T_H)$ is defined on the boundary of
the horizon. The volume bounded by the apparent horizon can be
calculated as
\begin{equation}\label{36}
V=\frac{4}{3}\pi\bigg[\frac{3\zeta_{0}t+3\mu t\psi-2\psi\Phi
_{1}}{2\psi}\bigg]^{3}.
\end{equation}
Using Eqs.\eqref{36} and \eqref{35}, we obtain
\begin{eqnarray}\nonumber
\dot{S}_{\text{in}}&=&\frac{1}{\psi \big(1-\frac{3 \zeta _0+3 \mu
\psi }{4 \psi }\big)}\bigg[\pi  \big(-2 \Phi_1 \psi +3 \zeta _0 t+3
\mu t \psi \big) \big(\frac{1}{\psi ^3}\big[\pi  \big(3 \zeta _0+3
\mu \psi \big){}^2 \big(-2 \Phi_1 \psi\\\nonumber &+&3 \zeta _0 t+3
\mu t \psi \big) \big(-\frac{12 (\mu -1) \psi ^3}{\big(3 t
\big(\zeta _0+\mu \psi \big)-2 \Phi_1 \psi \big){}^2}-\frac{12 \zeta
_0 \psi ^2}{\big(-2 \Phi_1 \psi +3 \zeta _0 t+3 \mu  t \psi
\big){}^2}\big)\big]\\\nonumber &-&\frac{12 \pi \big(3 \zeta _0+3
\mu \psi \big){}^2 \big(-2 \Phi_1 \psi +3 \zeta _0 t+3 \mu t \psi
\big)}{\big(3 t \big(\zeta _0+\mu \psi \big)-2 \Phi_1 \psi
\big){}^2}+\big(72 \pi \big(\zeta _0+\mu \psi \big) \big(3 \zeta
_0+3 \mu  \psi \big)\big)\\\nonumber &\times&\frac{\big(-2 \Phi_1
\psi +3 \zeta _0 t+3 \mu t \psi \big){}^2}{\big(3 t \big(\zeta
_0+\mu \psi \big)-2 \Phi_1 \psi \big){}^3}-\frac{108 \pi \big(\zeta
_0+\mu \psi \big){}^2 \big(-2 \Phi_1 \psi +3 \zeta _0 t+3 \mu  t
\psi \big){}^3}{\big(3 t \big(\zeta _0+\mu  \psi \big)-2 \Phi_1 \psi
\big){}^4}\\\nonumber &+&\big(\pi (3 \text{$\zeta $1}+3 \mu  \psi )
\big(-2 \Phi_1 \psi +3 \text{$\zeta $1} t+3 \mu  t \psi \big){}^2
\big(\frac{72 (\mu -1) \psi ^3 (\text{$\zeta $1}+\mu \psi )}{\big(3
t (\text{$\zeta $1}+\mu  \psi )-2 \Phi_1 \psi \big){}^3}\\\label{37}
&+&\frac{24 \text{$\zeta $1} \psi ^2 (3 \text{$\zeta $1}+3 \mu  \psi
)}{\big(-2 \Phi_1 \psi +3 \text{$\zeta $1} t+3 \mu  t \psi
\big){}^3}\big)\big)\frac{1}{2 \psi ^3}\big)\bigg].
\end{eqnarray}

The Hawking temperature is calculated as
\begin{eqnarray}\label{38}
T_{H}=\frac{1}{{2\pi
R_H}}\bigg({1-\frac{\dot{R_H}}{2HR_H}}\bigg)=\frac{\psi
\big(1-\frac{3\zeta _{0}+3\mu \psi}{4\psi}\big)}{\pi\big(-2Phi_{1}
\psi+3\zeta_{0}t+3\mu t\psi\big)}.
\end{eqnarray}
Considering the positive entropy ($S_{\text{in}}>0$) of matter
within the horizon, the universe total entropy should consistently
increase to adhere to the second law of thermodynamics
\begin{eqnarray}\label{39}
\dot{S}_{\text{tot}}&=&\dot{S}_{\text{on}}+\dot{S}_{\text{in}}
\\\nonumber&=&\frac{\pi K_b
\big(3\zeta_{0}+3\mu\psi\big)\big(-2\Phi_{1}\psi+3\zeta_{0}t+3\mu t
\psi\big)}{2\psi^{2}L^{2}_{p}}+\big(\pi\big(3\zeta_{0}t-2
\\\nonumber&\times&\Phi_{1}\psi+3\mu t\psi\big)\big(\frac{1}{2\psi^{3}}
\big(\pi\big(3\zeta_{0}+3\mu\psi\big)\big(3\zeta_{0}t-2 \Phi_1
\psi+3\mu t\psi\big)^{2}
\\\nonumber&\times&\big(\frac{24
\zeta_{0}\psi^{2}(3\zeta_{0}+3\mu\psi)}{\big(-2\Phi_{1}\psi+3\zeta
_{0}t+3\mu t\psi\big)^{3}}+\frac{72(\mu-1)\psi^{3} \big(\zeta
_{0}+\mu\psi\big)}{\big(3t\big(\zeta_{0}+\mu \psi\big)-2\Phi_{1}
\psi\big)^{3}}\big)\big)\\\nonumber&+&\frac{72\pi\big(\zeta_{0}+\mu
\psi\big)\big({3}\zeta _{0}+3\mu \psi\big)\big(-2\Phi_{1}\psi+3
\zeta_{0}t+3\mu t\psi\big)^{2}}{\big(3t\big(\zeta_{0}+\mu\psi
\big)-2 \Phi_1 \psi \big){}^3}\\\nonumber &-&\frac{108 \pi
\big(\zeta _0+\mu \psi \big){}^2 \big(-2 \Phi_1 \psi +3 \zeta _0 t+3
\mu  t\psi \big){}^3}{\big(3 t \big(\zeta _0+\mu \psi \big)-2 \Phi_1
\psi\big){}^4}+\big(\pi \frac{1}{\psi ^3}\big(3 \zeta _0\\\nonumber
&+&3\mu \psi \big){}^2 \big(\frac{-12 (\mu -1) \psi ^3}{\big(3 t
\big(\zeta _0+\mu \psi \big)-2 \Phi_1 \psi \big){}^2}-\frac{12 \zeta
_0 \psi ^2}{\big(3 \zeta _0 t-2 \Phi_1 \psi +3 \mu  t \psi
\big){}^2}\big)\\\nonumber &\times& \big(-2 \Phi_1 \psi +3 \mu  t
\psi +3 \text{t$\zeta $}_0\big)\big)-12 \pi \big(3 \zeta _0+3 \mu
\psi \big){}^2\\\label{40} &\times&\frac{ \big(-2 \Phi_1 \psi +3 \mu
t \psi +3 \text{t$\zeta $}_0\big)}{\big(3 t \big(\zeta _0+\mu \psi
\big)-2 \Phi_1 \psi \big){}^2}\big)\big)\frac{1}{\psi \big(1-\frac{3
\zeta _0+3 \mu \psi }{4 \psi }\big)}\geq 0.
\end{eqnarray}
Figure \textbf{6} demonstrates that the overall cosmic entropy rises
with time, illustrating the increasing Hawking temperature after the
bounce period. This behavior shows the entropy production
compatibility and the fundamental rules of cosmic thermodynamics.
Moreover, the growing Hawking temperature indicates the system
advancement heading to an equilibrium state throughout the expansion
phase.  his rise in temperature matches theoretical data for the
phenomena involving reheating processes following a cosmic bounce.
\begin{figure}
\epsfig{file=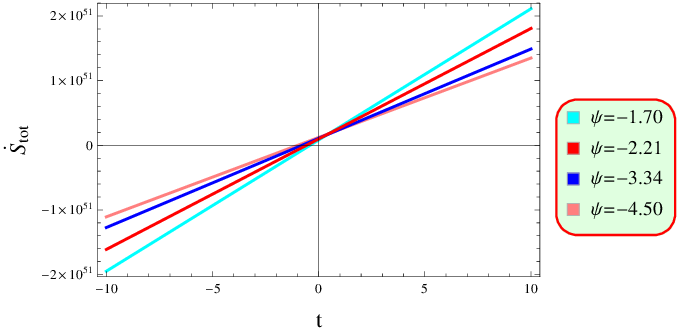,width=.5\linewidth}
\epsfig{file=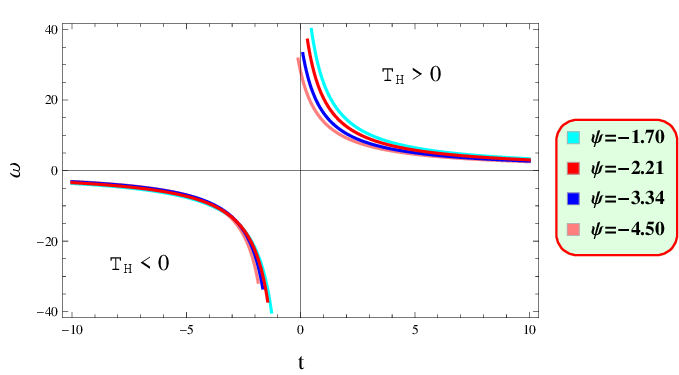,width=.5\linewidth}\caption{Plots of Entropy
and Hawking Temperature}
\end{figure}

\section{Discussions and Final Outcomes}

The fundamental cosmological framework has captured the significance
in elucidating the dynamics of the cosmos, but it faces notable
discrepancies, especially the appearance of the singularities where
the curvature of spacetime becomes infinite and the breakdown of the
physical laws.  To take into consideration these constraints, a
different framework such as bouncing cosmology offers promising
cosmic scenario that presents pre-expansion contraction cosmic
phases, hence prevents the primordial singularities. From this
perspective, a bounce offers a smooth and transition between
multiple cosmic stages that may not be singular. Bulk viscosity, a
method that regulates variations but additionally integrates a
cosmic history that is more plausible and realistic, is one
effective technique for such a bounce.  Viscosity in bulk has a
major impact on the cosmic properties and broadens the ideal fluid
premises.  By examining the distribution of viscous matter, we look
into the consequences of cosmic behaviors and bulk viscosity in the
$f(\mathcal{Q})$ theory.

The principal application of our modified model lies in its ability
to resolve the initial singularity problem through the realization
of a non-singular bouncing cosmology. As opposed to inflationary
models, which bounce backwards to the big bang singularity. Bouncing
cosmology offers a compelling alternative where the universe
undergoes a smooth transition from a contracting phase to an growing
one.  Our framework is based on the $f(\mathcal{Q})$ gravity
augmented with bulk viscosity, providing a geometrically driven
technique for this bounce that does not require unusual matter
fields or high-energy corrections from quantum gravity. Furthermore,
we can successfully simulate dissipative processes in the early
universe by integrating bulk viscous effects into the perfect fluid.
This makes our model directly applicable to the study of
cosmological dynamics, which is especially important for
comprehending the onset of the arrow of time in bouncing scenarios.
Additionally, the model serves as a theoretical testing ground for
geometric entropy production, helping bridge gravitational theory
with thermodynamic irreversibility, which is a growing topic of
interest in cosmology.

Our main findings are summed up here.
\begin{itemize}
\item
The Hubble analysis successfully captures the pre-bounce cosmic
contraction, transition at the bouncing epoch ($H=0$) and
post-bounce expansion for negative values of $\psi$ (right panel of
Figure \textbf{1}). The temporal derivative of the Hubble parameter
highlights the smooth transition from contraction to expansion
phases of the cosmic bounce (left panel of Figure \textbf{1}).
\item
The deceleration variable serves an indicator of the model's
consistency with the universe expansion evolution inferred from
observed cosmic behavior. Specifically, the negative profile before
and after the bounce, signifies the accelerated cosmic expansion
(Figure \textbf{2}).
\item
The positive trait of energy density and negative pressure curve
signifies the accelerated cosmic expansion. This unique matter
variables behavior supports the realistic non-singular bouncing
cosmology (Figure \textbf{3}).
\item
The ECs and EoS parameter reflects a transition from an early
stiff-matter dominated regime to an expansion period. The
incremental decay towards zero illustrates that the universe
convergently approaches to a radiation dominated or dust-like
configuration expansion at late epochs (Figures \textbf{4} and
\textbf{5}).
\item
The rise in entropy production about the bouncing epoch ensures
cosmic acceleration and compatibles with the generalized second law
of thermodynamics, signifying a smooth modulation between
contracting and expanding cycles (left plot of Figure \textbf{6}).
The intensifying Hawking temperature signifies equilibration in the
expanding period, underscoring the role of bulk viscosity in
sustaining a thermodynamic bouncing cosmology while eluding the
singularities (right plot of Figure \textbf{6}).
\end{itemize}

The $f(\mathcal{Q})$ gravity framework generalizes the teleparallel
formulation, where the gravitational interaction is governed not by
curvature or torsion, but by the non-metricity scalar. This geometry
permits richer cosmological behavior while allowing us to
investigate gravitational dynamics that preserves second-order field
equations. Our research makes use of this adaptability to examine
dissipative processes through bulk viscosity, which occurs
spontaneously in realistic cosmological fluids. A generalized form
of the bulk viscosity coefficient is presented. This form
encompasses a wide range of dissipative behaviors and is motivated
by past research in viscous cosmology. In order to maintain
analytical tractability, we have chosen a straightforward but
non-trivial polynomial form of $f(\mathcal{Q})$. This keeps the
model from becoming overly complicated while allowing for meaningful
physical interpretation. In order to evaluate the bounce physical
feasibility, the second law of thermodynamics must be investigated
by calculating the entropy production rate. It is an essential test
to confirm whether the suggested bounce scenarios adhere to basic
physical laws, not just an additional analysis. Even though the
model has multiple components, each one is essential to address a
particular facet of cosmic evolution and singularity resolution.
Additionally, our research offers a novel combination of
thermodynamics, viscous fluid dynamics and non-metric gravity, an
interaction that is rarely discussed in the literature.

In our viscous $f(\mathcal{Q})$ cosmological framework, the
thermodynamic results show a high degree of agreement with similar
findings from recent black hole studies conducted in modified
gravity and quantum corrected contexts.  For example, under the
influence of quantum gravity, the adjusted Hawking temperature and
entropy corrections for a Dyonic black hole encircled by a perfect
fluid. The stability and equilibrium convergence for the horizon of
large radii were confirmed by the observation that quantum
corrections increase the logarithmic entropy production and decrease
the Hawking temperature. In our framework, a similar stabilization
profile is visible, with the Hawking temperature approaching a
constant optimal value during the expansion and the entropy rising
readily near the bounce epoch. This suggests a smooth modulation
towards thermodynamic equilibrium rather than rapid heating.
Similarly, the entropy for a modified Schwarzschild-Rindler black
hole and corrected temperature dynamics has been examined in
\cite{103}, confirming that quantum variables limit the extreme
temperature rise, specifically reflecting the viscous damping effect
in our cosmic framework. Moreover, our findings display a fine
mapping with \cite{104},where $(\lambda)$ and $(\mu)$ were found to
persuade the Hawking temperature and effective potential,
influencing energy radiation. Correspondingly, in our framework, the
bulk viscosity variable $(\zeta)$ and the chosen $f(\mathcal{Q})$
function play as effective couplings, leading the cosmic
thermodynamic evolution. The behavior of the EoS variable in our
study modulations between the phantom ($\omega<-1$) era and
quintessence $(-1<\omega<-\frac{1}{3})$ period, aligning with the
phenomenological domains reported in \cite{104} for the ModMax-AdS
scheme, where the non-linear field mapping give similar modulations.

The $f(\mathcal{Q})$ model exhibits stable characteristics and
consistently corresponds with the present cosmic expansion. It is
important to note that our findings are aligns with the existing
observational data \cite{105}. The data was obtained from diverse
observational approaches with a confidence level of 95\%. To connect
our theoretical findings with observations, we compare our results
for the EoS parameter with the latest constraints from the Planck
collaboration \cite{105}.  Additionally, our findings agree with
recent theoretical and observational \cite{108}. Our investigation
demonstrates that the inclusion of bulk viscosity contributes to the
stabilization of pressure and energy density, thereby facilitating a
physically viable bounce scenario. Furthermore, the analysis of
cosmographic parameters indicate that the model prediction of the
accelerated universe is influenced by DE-like behavior. This model
is entirely geometric without including realistic matter
configuration. On the other hand, our study focussed on introducing
the bulk viscosity via a new parametrization, facilitating the
modeling of dissipative phenomena through early cosmos. Thus, our
work is observationally relevant, more comprehensive and physically
realistic, offering a clear exploration in the study of cosmic
bounce models.

We envision several promising avenues in which this model can be
extended or applied in future work. An important next step is to
investigate cosmological perturbations in this framework. While the
current work is theoretical in nature, the model can be calibrated
using background cosmological data. We hope to explore loop quantum
gravity-inspired corrections or effective semi-classical models that
build upon this classical foundation. This not only situates our
work within the broader context of modern cosmology, but also
highlights its adaptability and relevance for future research.
\\\\
\textbf{Data Availability Statement:} No new data is generated in
this study.

\end{document}